\documentclass{article}
\usepackage{graphicx}
\usepackage{citesort}
\usepackage{a4wide}
\usepackage{epsf}
\usepackage{bm}

\newcommand{\la}{\left\langle}
\newcommand{\ra}{\right\rangle}

\newcommand{\bx}{{\bm x}}
\newcommand{\bu}{{\bm u}}
\newcommand{\br}{{\bm r}}

\newcommand{\ba}{{\bm a}}

\title{Joint statistics of acceleration and vorticity in fully
  developed turbulence}
\author{Luca Biferale$^{1}$ and Federico Toschi$^{2}$\\
  $^{1}$Dipartimento di Fisica, Universit\`a degli Studi di Roma ``Tor Vergata''\\
  and INFN, Via della Ricerca Scientifica 1, I-00133 Roma, Italy\\
  $^{2}$Istituto per le Applicazioni del Calcolo, CNR, \\
  Viale del Policlinico 137, I-00161 Roma, Italy}
\begin{document}
\maketitle

\begin{abstract}
  We report results from a high resolution numerical study of fluid
  particles transported by a fully developed turbulent flow at
  $R_{\lambda} =280$.  Single particle trajectories were followed for
  a time range spanning more than three decades, from less than a
  tenth of the Kolmogorov time-scale up to one large-eddy turnover
  time. We present results concerning acceleration statistics and the
  statistics of trapping by vortex filaments conditioned to the local
  values of vorticity and enstrophy. We distinguish two different
  behaviors between the joint statistics of vorticity and centripetal
  acceleration or vorticity and longitudinal acceleration.
\end{abstract}

Lagrangian statistic of particles advected by a turbulent velocity
field, $\bm u(\bm x,t)$, is important both for theoretical
implications \cite{K65} and for applications, such as the development
of phenomenological and stochastic models for turbulent mixing
\cite{pope,aringazin}. Recently, important  advances in experimental
techniques for measuring Lagrangian turbulent statistics
\cite{cornell,pinton,ott_mann,leveque,bodeprl} have been achieved.
Direct numerical
simulations (DNS) also offer very high accuracy albeit at a slightly
lower Reynolds number \cite{yeung,BS02,IK02,borgas}.  We analyze
Lagrangian data obtained from a recent DNS
of forced homogeneous isotropic turbulence
\cite{biferale04,biferale04b}, performed on $512^3$ and $1024^3$ cubic
lattices with Reynolds numbers up to $R_\lambda \sim 280$. The
Navier-Stokes equations were integrated using fully de-aliased
pseudo-spectral methods for a total time $T \approx T_L$.  Two
millions of Lagrangian particles (passive tracers) were injected into
the flow once a statistically stationary velocity field had been
obtained. The positions and velocities of the particles were stored at
a sampling rate of $0.07 \tau_\eta$. The velocity of the Lagrangian
particles was obtained using linear interpolation of the Eulerian
field.  
\begin{figure}[!b]
  \hbox{ \includegraphics[draft=false,width=0.50\hsize]{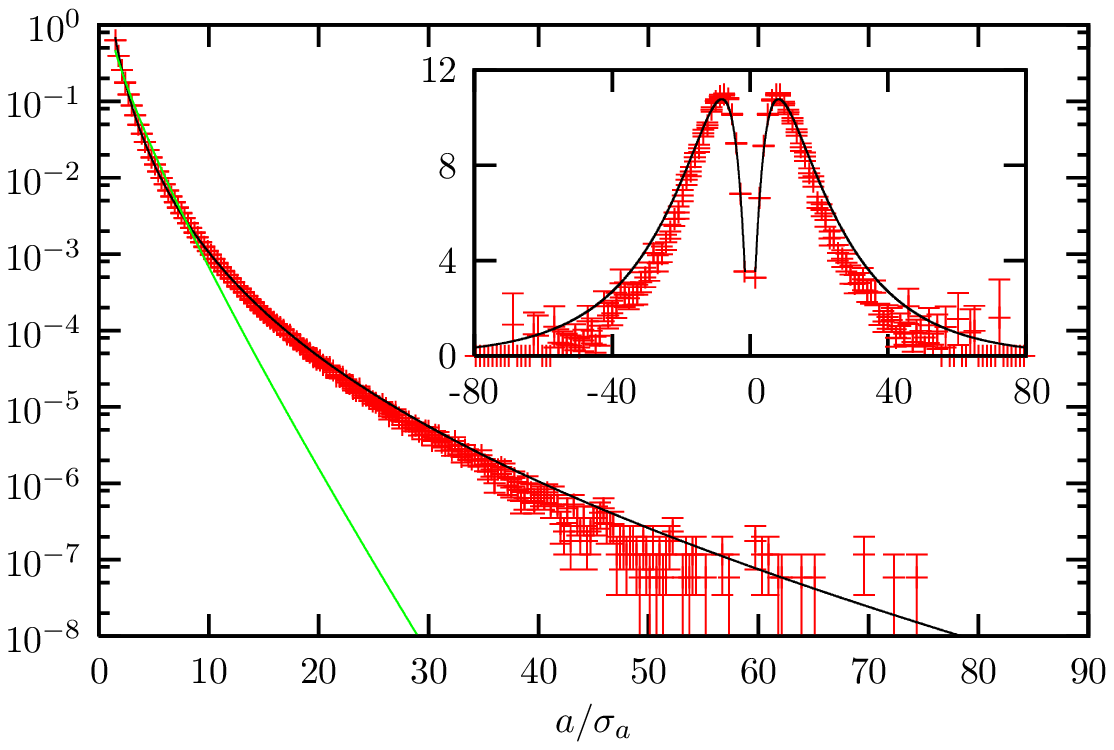}
    \includegraphics[draft=false,width=0.50\hsize]{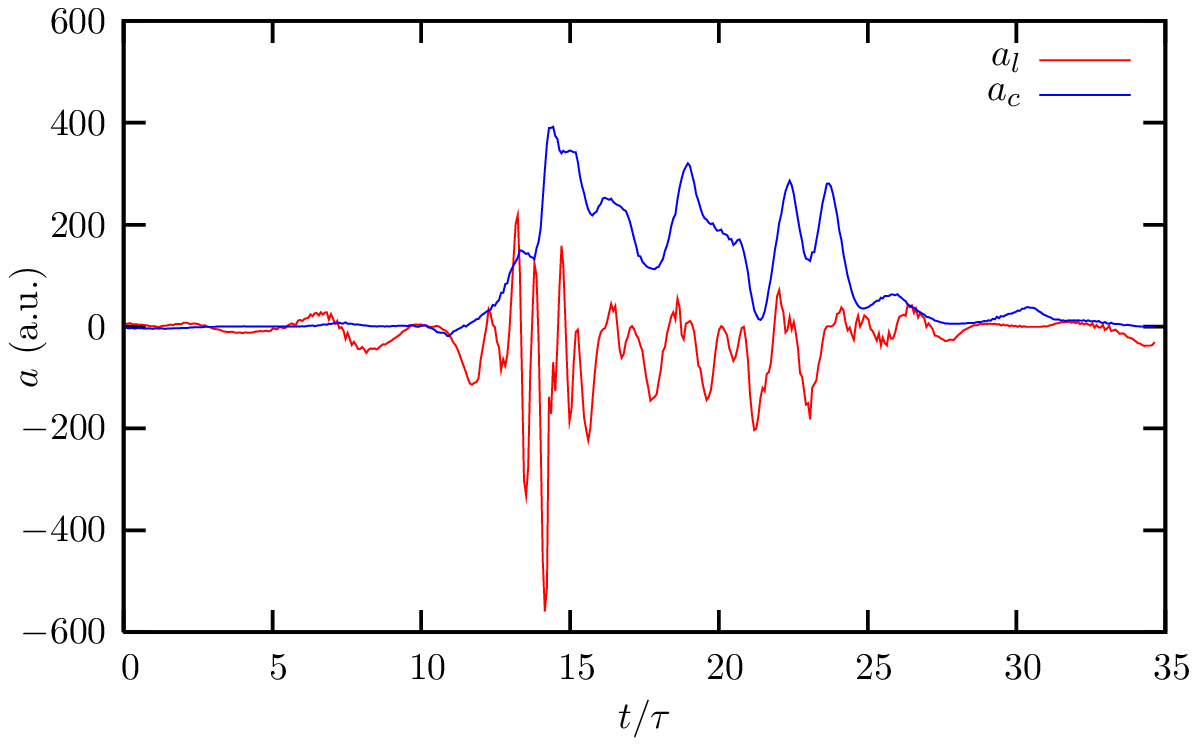} }
\begin{center}
\caption{
  (Left panel).  Log-linear plot of the acceleration pdf, ${\cal
    P}(a)$.  The crosses are the DNS data, the solid black line is the
  multifractal prediction \cite{biferale04b} and the green line is the
  K41 prediction. The statistical uncertainty in the pdf is quantified
  by assuming that fluctuations grow proportional to the square root
  of the number of events.  Inset: we show $a^4 {\cal P}(a)$ to check
  the statistical convergence up to fourth order moments, the
  continuous line is the multifractal prediction presented in
  \cite{biferale04b}.  (Right panel). We show, in natural units, the
  behavior of one component of the centripetal and of the longitudinal
  acceleration. Notice the strong sign persistence of the centripetal
  acceleration with respect to the longitudinal one.
\label{fig1}}
\end{center}

\end{figure}
Acceleration was measured both as the derivative of the particle
velocity and by direct computation from all three forces acting on the
particle (i.e.  pressure gradients, viscous forces and large scale
forcing): the two measurements were found to be in very good agreement
(with the latter being less noisy).  Finally, the flow was forced by
keeping the total energy constant in each of the first two wavenumber
shells.  For more details on the simulation, see
\cite{biferale04,biferale04b,biferale04c}.  Recently much attention
focused on the statistics of acceleration with and without
conditioning on the local structure of the flow
\cite{cornell,bodeprl,biferale04b,biferale04c,jfp}. Some
phenomenological description of such statistics using multifractal
\cite{biferale04b,arimitsu} or quasi-equilibrium distribution have
been also proposed \cite{beck} (for a critical summary of these
attempts see \cite{GK03}).  In this paper we concentrate mainly on the
statistics of trapping events, i.e.  those cases when the particle is
captured inside a vortex filaments for a time lag considerably larger
than the Kolmogorov eddy turn over time, $\tau_{\eta}$
\cite{cornell,biferale04}.  These events contribute to the statistics
of the particle acceleration, ${\cal P}(a)$, with extremely intense
values, up to $80$ times the root mean squares acceleration, at the
Reynolds number here investigated (see Fig.~\ref{fig1}). In previous
analysis \cite{biferale04,biferale04b,biferale04c} we have shown that
the trapping in vortex filaments is responsible for a strong
deterioration of scaling properties of Lagrangian Structure functions
for time lags of the order of $\tau_{\eta}$. In a previous publication
\cite{biferale04c} we shown that trapping into vortex filaments leads
to very different dynamical behavior for centripetal and longitudinal
acceleration (see also \cite{jfp}). Being the latter highly
oscillating in time, while the former almost constant and with very
high amplitude (see Fig.~\ref{fig1}).  The big difference between the
dynamical properties of longitudinal and centripetal acceleration is
already detected by inspecting their temporal correlations:
\begin{equation}
C_c(\tau) = \la a^x_c(t) a^x_c(t +\tau)  \ra \;\;\;  C_l(\tau) = \la a^x_l(t) a^x_l(t +\tau) \ra
\end{equation}
where we focus only on one component of the acceleration correlation
for simplicity.  In Fig.~\ref{fig:corre} we present the DNS results
for these two quantities. As one can see, while the longitudinal
acceleration decreases with a characteristic time which is comparable
with the Kolmogorov time $\tau_{\eta}$, the centripetal one has a much
slower time decay. Indeed, as shown in the inset, the centripetal
correlation, $C_c(\tau)$, possess two different decays \cite{cornell,pinton,jfp}. The first one, for short time lags
(up to $4\div 5 \tau_{\eta}$) is comparable with the one of the
longitudinal correlation. The second decay, with a characteristic time
of the order of $10\div 15 \tau_{\eta}$ is established for longer time
lags. This second exponential tail is the signature of the strong
persistence inside vortex filaments.
\begin{figure}[ht]
\begin{center}
\includegraphics[draft=false,width=.9\hsize]{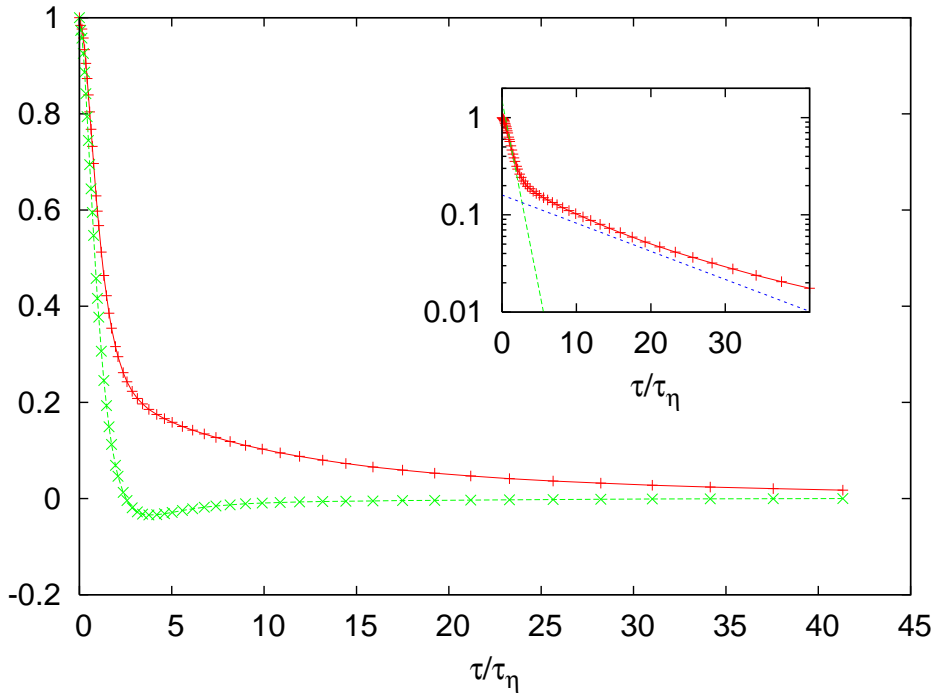}
\caption{Correlation functions of the centripetal ($+$, red curve) and
  longitudinal ($\star$, green curve) accelerations. Notice the much
  slower decay showed by the centripetal component. Inset: Log-linear
  plot of the centripetal correlation function, $C_c(\tau)$, with
  superposed the two exponential behavior for short time lags,
  $\exp(-\tau/(1.1 \tau_{\eta}))$ and for large time lags,
  $\exp(-\tau/(15 \tau_{\eta}))$, obtained with a best fit at short
  and long times respectively.}
\label{fig:corre}
\end{center}
\end{figure}
In order to better quantify this trapping events, we present in this
paper some  results on the statistical
properties of the longitudinal and centripetal acceleration
conditioned to the local properties of the vorticity field, ${\bm
  \omega}(\bx)$ and of the enstrophy, $\Omega(\bx)$.
The presence of persistent intense vorticity structure at dissipative
scales may have a strong feedback on the energy cascade mechanism. It
is not clear if and how these structures may also influence the
inertial range physics through non-local effects (filaments are quite
elongated in one direction). Other studies similar to the one
presented here, but focused on the joint probability distribution of
the energy transfer and the vorticity field may help in clarifying
this important issue.

\section{Joint statistics}
\begin{figure}[ht]
\begin{center}
\includegraphics[draft=false,width=.9\hsize]{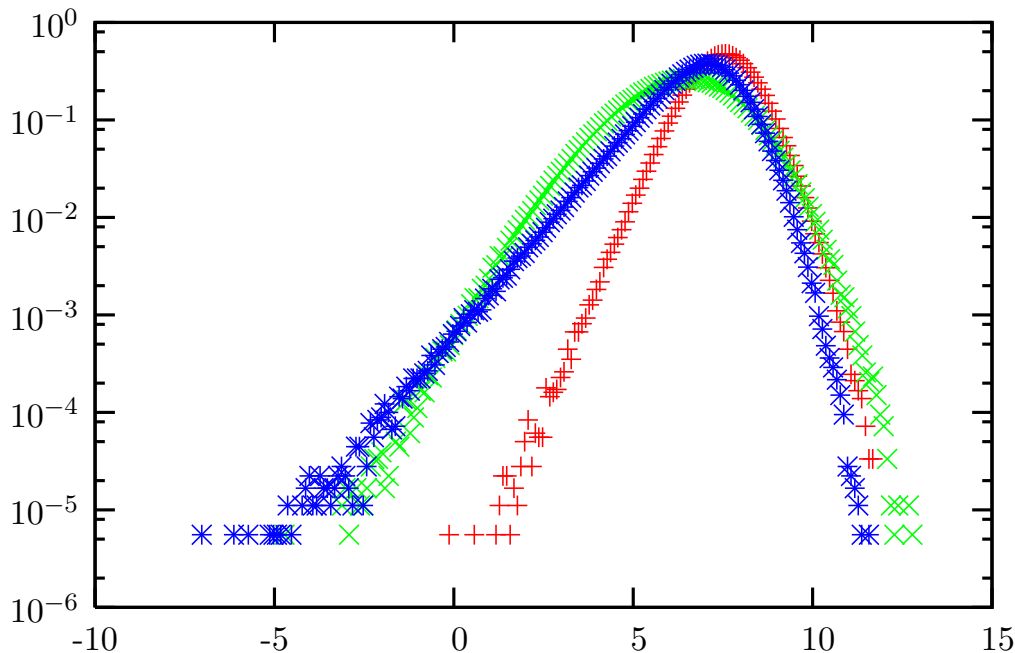}
\caption{
  Logarithm of the Probability density function of the instantaneous
  centripetal acceleration, ${\cal P}(\log(|\ba_c|/\eta))$ ($+$, red
  curve), longitudinal acceleration ${\cal P}(\log(|\ba_l|/\eta))$
  ($\star$, blue curve) and of the enstrophy, ${\cal P}(\log(\Omega))$
  ($\times$, green curve).  Acceleration amplitudes are rescaled with
  the Kolmogorov scale $\eta$ to test relation (\ref{eq:cen}).}
\label{fig:pdf}
\end{center}
\end{figure}

We want to study the correlation between the probability to observe a
large value in the acceleration fluctuations and the local Eulerian
structure of the flow. To do that we first define the longitudinal and
centripetal instantaneous acceleration to be
$${\bm a}_l \equiv ({\bm a \cdot \hat {\bm v}}){ \hat {\bm v}}$$
and
$${\bm a}_c \equiv {\bm a \times \hat {\bm v}}$$
respectively; where
$\hat {\bm v}(t)$ is the particle velocity unit vector,
 $ {\bm v}(t) = \bu(\br(t),t)$, and $\bu(\bx,t)$ is the
Eulerian velocity field.  Similarly, we are interested to the square
of the antisymmetric part of the stress tensor, the enstrophy,
$\Omega(\bx)$:
\begin{equation}
\Omega(\bx) = \frac{1}{2}\sum_{ij} (\partial_i v_j - \partial_j v_i)^2
=  {\bm \omega}^2
\end{equation}
where ${\bm \omega}$ is the vorticity. In a purely circular motion of
radius $r$, around a vortex filament with vorticity, ${\bm \omega}$,
we expect that the centripetal acceleration can be expressed as:
\begin{equation}
|\ba_c| \propto {\bm \omega}^2 r  \sim \Omega r
\label{eq:cen}
\end{equation}
which is a direct links between the local properties of the enstrophy
and the statistics of the centripetal acceleration. The above argument
does not fix the ${\cal O}(1)$ proportionality prefactor, which
depends on the exact shape of the vortex filament. In order to probe
how much the rare, but intense, events characterizing the tails of the
acceleration statistics are indeed caused by trapping in regions of
high vorticity we studied the joint probability densities of
centripetal acceleration and enstrophy, compared with the joint
probability density of longitudinal acceleration and enstrophy (for
recent similar investigation in experimental and numerical data, see
\cite{jfp}):
\begin{equation}
{\cal P}(\log(\ba_c),\log(\Omega)), \;\;\; {\cal P}(\log(\ba_l),\log(\Omega))
\label{eq:condpdf}
\end{equation}
where we concentrate directly with the logarithm of the variable to
 better appreciate extreme events.  Let us first show in
Fig.~\ref{fig:pdf} the probability density function of the three
--unconditioned-- quantities:
$${\cal P}(\log(\ba_c)),\;\; {\cal P}(\log(\Omega))),\;\; {\cal
  P}(\log(\ba_l)).$$
According to relation (\ref{eq:cen}) one would
expect that if intense tails in the centripetal acceleration are due
to the vortex trapping at Kolmogorov scale $\eta$, then the two PDF,
${\cal P}(\log(|\ba_c|))$ and ${\cal P}(\log(\Omega))$ should have the
same behavior for large values after the rescaling $|\ba_c|/\eta$ is
made.  This is what we showed in Fig.~\ref{fig:pdf} where one can
see that the two tails corresponding to ${\cal P}(\log(|\ba_c|/\eta))$
and to ${\cal P}(\log(\Omega))$ almost perfectly superpose ( within
${\cal O}(1)$ prefactors which are out of control with dimensional
arguments).

\begin{figure}[ht]
\begin{center}
  \includegraphics[draft=false,width=1.0\hsize]{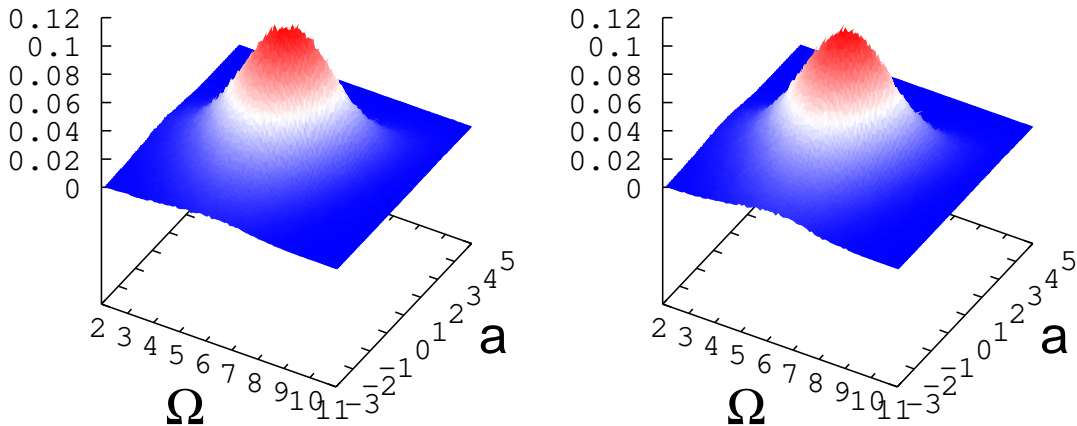}
\caption{Joint probability distribution function for one component of the 
  centripetal acceleration and enstrophy, ${\cal
    P}(\log(a^x_c),\log(\Omega))$ (Left panel), and of longitudinal
  acceleration and enstrophy ${\cal P}(\log(a^x_l),\log(\Omega))$
  (Right panel). Notice that the two shapes are almost
  indistinguishable.}
\label{fig:condpdf}
\end{center}
\end{figure}

However, from the unconditional PDF shown in Fig.~\ref{fig:pdf} there
are not strong signs distinguishing the dynamical properties of
centripetal and longitudinal acceleration.  Similarly, the joint
probability densities (\ref{eq:condpdf}) plotted in Fig.
\ref{fig:condpdf} do not show any quantitative differences between
the statistics of the centripetal and longitudinal acceleration.
This is quite natural, because the motion of a particle in a turbulent
field will be characterized by different accelerations and
decelerations, not necessarily associated with spiraling motion (on
average the mean value of the acceleration will be zero). The
distinguishing character of vortex trapping is the strong persistence
of the direction -and amplitude- of the centripetal acceleration for
time lags much larger than $\tau_{\eta}$, at variance with what
happens to the longitudinal acceleration, see right panel of Fig.
\ref{fig1}.

To make this statement quantitative, we have studied the running
average of the centripetal and longitudinal acceleration, over a time
window varying up to $10\tau_{\eta}$ \cite{biferale04c}:
\begin{eqnarray}
 {\bm a}^{\Delta}_c(t) \equiv   {1\over
\Delta}\int_{t-\Delta/2}^{t+\Delta/2} dt' {\bm a}_c(t');\\ 
{\bm a}^{\Delta}_l(t) \equiv  {1\over \Delta}\int_{t-\Delta/2}^{t+\Delta/2} dt' {\bm a}_l(t').
\label{eq:mean}
\end{eqnarray}
We expect that the pdfs of the averaged centripetal and longitudinal
acceleration will behave very differently at increasing the window
size, $\Delta$.  In particular, the strong persistence of the
centripetal acceleration up to $10 \tau_{\eta}$ suggests that the
joint PDF of centripetal acceleration and enstrophy, ${\cal P}(
{\log(\bm a}^{\Delta}_c),\log(\Omega))$ should remain almost unchanged
at varying $\Delta$, while the longitudinal one ${\cal P}( \log({\bm
  a}^{\Delta}_l),\log(\Omega))$ should experience a strong depletion,
of events with simultaneous intense values of acceleration and
vorticity.
\begin{figure}[ht]
\begin{center}
  \includegraphics[draft=false,width=1.0\hsize]{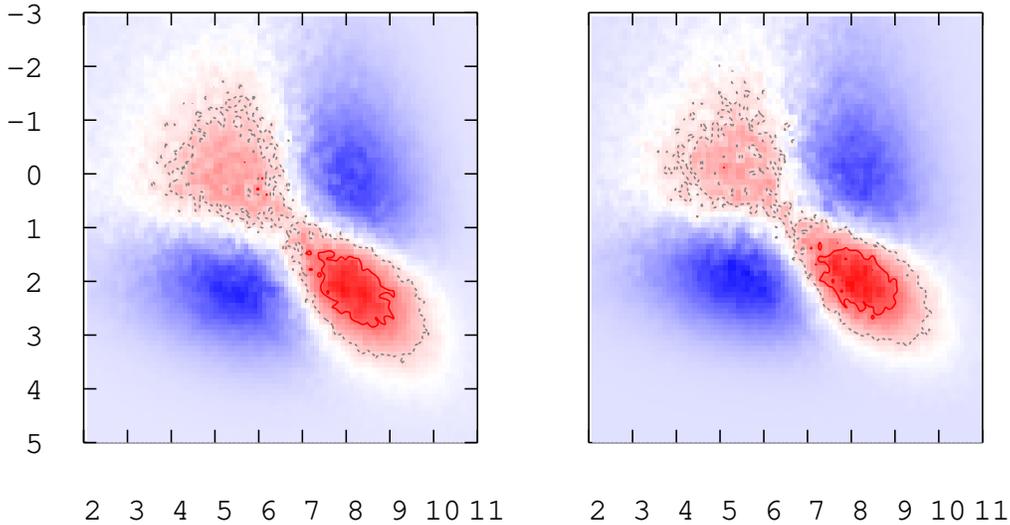}
\caption{
  (Left panel). Residual value, ${\cal
    R}(\log(\ba^{\Delta}_c),\log(\Omega))$ for the centripetal
  acceleration and the local enstrophy without any average in time
  (i.e. $\Delta=0$). Acceleration is plotted on the $y$ axis while the
  enstrophy is plotted on the $x$ axis. (Right panel). The same but
  for the longitudinal acceleration, ${\cal
    R}(\log(\ba^{\Delta}_l),\log(\Omega))$. The isoline is drawn for a
  fixed value of ${\cal R}$, to guide the eyes.  Notice the strong
  correlation between intense vorticity events and intense centripetal
  and longitudinal acceleration (bottom right corner).  Without
  averaging in time, no clear distinction between longitudinal and
  centripetal signal can be detected. As expected, the distinction
  arises only when persistent effects are investigated, see Figure
  (\ref{fig:resave100}).}
\label{fig:resave1}
\end{center}
\end{figure}

\begin{figure}[ht]
  \includegraphics[draft=false,width=1.0\hsize]{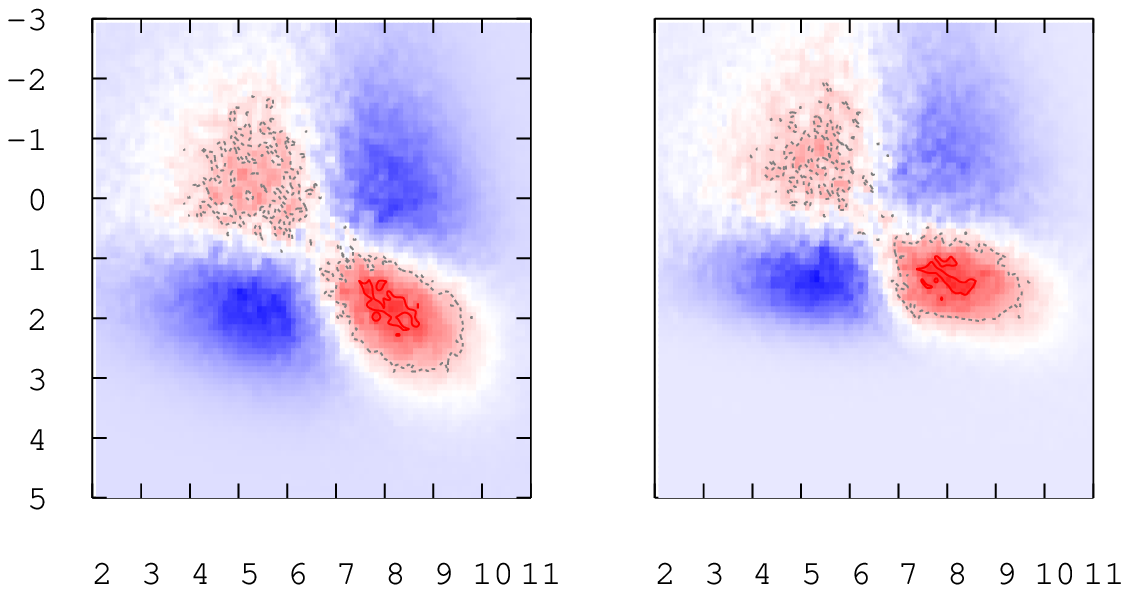}
\caption{
  (Left panel). Residual value, ${\cal
    R}(\log(\ba^{\Delta}_c),\log(\Omega))$, for the centripetal
  acceleration and the local enstrophy after averaging in time, over a
  window $\Delta= 9 \tau_{\eta}$. (Right panel). The same but for the
  longitudinal acceleration, ${\cal
    R}(\log(\ba^{\Delta}_l),\log(\Omega))$. The isoline is drawn for a
  fixed value of ${\cal R}$, to guide the eyes.  Notice the
  persistence of the strong correlation present between intense
  vorticity events and intense centripetal acceleration. On the other
  hand, now the longitudinal acceleration is not any more correlated
  for intense events with the enstrophy (compare the bottom right
  corners of both panels).}
\label{fig:resave100}
\end{figure}
  A way to offer a quantitative
measurement of the correlation between the two variables entering the
joint PDF, ${\cal P}(x,y)$, is to plot the residual value, ${\cal
  R}(x,y) \equiv {\cal P}(x,y) - {\cal P}(x) {\cal P}(y)$, obtained by
subtracting the probability density in the case of fully uncorrelated
variables, ${\cal P}(x) {\cal P}(y)$.  This is what we show in Figs.
(\ref{fig:resave1}-\ref{fig:resave100}) where we compare the residual
value for the averaged centripetal acceleration and the local
enstrophy:
\begin{equation}
{\cal R}(\log(\ba^{\Delta}_c),\log(\Omega)) \equiv {\cal P}(\log(\ba^{\Delta}_c),\log(\Omega)) -{\cal P}(\log(\ba^{\Delta}_c))  {\cal P}(\log(\Omega)) 
\label{eq:diff1}
\end{equation}
and the same quantity for the longitudinal one:
\begin{equation}
{\cal R}(\log(\ba^{\Delta}_l),\log(\Omega)) \equiv {\cal P}(\log(\ba^{\Delta}_l),\log(\Omega)) -{\cal P}(\log(\ba^{\Delta}_l))  {\cal P}(\log(\Omega)) 
\label{eq:diff2}
\end{equation}
for two different values of the window, $\Delta$.  Let us stress that
the difference from zero of the residual values here defined give a
direct measurement of the level of correlation between the two
fluctuating quantities.  Positive values of the difference in
(\ref{eq:diff1}) and (\ref{eq:diff2}) indicate correlation while
negative values indicate anti-correlation.  In Fig.~\ref{fig:resave1}
we plot the case of (no-average in time) $\Delta=0$, and in
Fig.~\ref{fig:resave100} the case of $\Delta= 9 \tau_{\eta}$. As one
can see, in the first case, (Fig.~\ref{fig:resave1}), a strong
correlation between extreme events in the acceleration and the
vorticity is observed, although almost no difference is detectable
between the longitudinal and centripetal acceleration (right and left
panels, respectively).  On the other hand, once averaged in time, i.e.
filtering out the high oscillation in the longitudinal acceleration, a
strong pick in the acceleration-vorticity contour plot is visible only
for the centripetal case (see left panel of Fig.~\ref{fig:resave100}.
This is in our view a simple but effective way to highlight the strong
correlation between the observed high tails in the acceleration
distribution and the presence of vortex filaments in the Eulerian
field.

\section{Conclusions}
We have presented results on the joint Lagrangian acceleration
statistics and Eulerian vorticity field from DNS of fully developed
turbulence.  In particular we have shown the existence of strong
correlation between intense and persistent centripetal acceleration
and the presence of intense vortical structure. No effects similar to
this one is detected for the longitudinal components of the
acceleration. We interpret this as an evidence of particle trapping in
vortex filaments. Further investigations are  necessary in order to
understand if and how this events affects the local energy transfer
mechanism and, if it is  the case, 
how to include  the presence of these filamentary
structure in the  energy cascade phenomenology.

We thank G. Boffetta, A. Celani, B. Devenish and A.  Lanotte for
discussions.  We thank the supercomputing center CINECA (Bologna,
Italy) and ``Centro Ricerche e Studi Enrico Fermi'' for the resources
allocated for this project. We acknowledge Dr. Nazario Tantalo for
technical assistance.


\section*{References}
\begin{thebibliography}{99}
\bibitem{K65} Kraichnan RH 1965 {\em Phys. Fluids} {\bf 8} 575.
\bibitem{pope} Pope SB 2000 {\it Turbulent Flows} (Cambridge
  University Press, Cambridge).
\bibitem{aringazin} Aringazin AK and Mazhitov MI 2003 {\em Phys. Lett.
    A} {\bf 313} 284.  Aringazin AK and Mazhitov MI 2004 {\em Phys.
    Rev. E} {\bf 70}, 036301.
\bibitem{cornell} La Porta A, Voth GA, Crawford AM, Alexander J and
  Bodenshatz E 2001 {\em Nature} {\bf 409} 1017.  Voth GA {\it et al}
  2002 {\em J. Fluid Mech.} {\bf 469} 121.  Mordant N {\it et al.}
  2003 {\em Physica D} {\bf 193} 245.
\bibitem{pinton} Mordant N {\it et al.} 2003 {\em J. Stat. Phys.} {\bf
    113} 701. Mordant N {\it et al.} 2002 {\em Phys. Rev. Lett.}  {\bf
    89} 254502.  Mordant N {\it et al.} 2001 {\em Phys. Rev.  Lett.}
  {\bf 87} 214501.
\bibitem{ott_mann} Ott S and Mann J 2000 {\em J. Fluid Mech.} {\bf
    422} 207.
\bibitem{leveque} Chevillard L, Roux SG, Leveque E {\it et al.}  2003
  {\em Phys. Rev. Lett.} {\bf 91} 214502.
\bibitem{bodeprl} A. M. Crawford, N. Mordant and E. Bodenschatz 2005
  {\em Phys.  Rev. Lett.} {\bf 94}, 024501.
\bibitem{yeung} Yeung PK 2002 {\em Ann. Rev. Fluid Mech.} {\bf 34}
  115.  Yeung PK 2001 {\em J. Fluid Mech.} {\bf 427} 241.  Vedula P
  and Yeung PK 1999 {\em Phys. Fluids} {\bf 11} 1208.
\bibitem{borgas} Yeung PK and Borgas M 2004 {\em J. Fluid Mech.} {\bf
    503} 93.
\bibitem{BS02} Boffetta G. and Sokolov IM 2002 {\em Phys. Rev. Lett.}
  {\bf 88} 094501.
\bibitem{IK02} Ishihara T and Kaneda Y 2002 {\em Phys. Fluids} {\bf
    14} L69.


\bibitem{beck} Beck C 2003 {\em Europhys. Lett.} {\bf 64} 151.  Beck C
  2001 {\em Phys. Lett. A} {\bf 27} 240.
\bibitem{arimitsu} Arimitsu T and Arimitsu N 2003 {\em Physica D} {\bf
    193} 218.
\bibitem{GK03} Gotoh T and Kraichnan RH 2004 {\em Physica D} {\bf 193}
  231.
\bibitem{jfp} Mordant N, Leveque E and Pinton JF 2004 {\em New J.
    Phys.} {\bf 6} 116.
\bibitem{biferale04} Biferale L, Boffetta G, Celani A, Lanotte A and
  Toschi F 2005 {\em Phys. Fluids} {\bf 17} 021701.
\bibitem{biferale04b} Biferale L, Boffetta G, Celani A, Devenish BJ,
  Lanotte A and Toschi F 2004 {\em Phys. Rev. Lett.} {\bf 93} 064502.
\bibitem{biferale04c} Biferale L, Boffetta G, Celani A, Devenish BJ,
  Lanotte A and Toschi F 2004 {\em Jour. of Turbulence} to appear
  (2005); lanl.arXiv.org nlin.CD/0501041
\end{thebibliography}
\end{document}